\documentclass[a4paper]{article}

\usepackage{INTERSPEECH2020}

\usepackage{xcolor}
\usepackage{hyperref}
\usepackage{multirow,booktabs,colortbl,tabularx}
\usepackage{amsmath}

\title{Vocoder-Based Speech Synthesis from Silent Videos}
\name{Daniel Michelsanti$^1$, Olga Slizovskaia$^2$, Gloria Haro$^2$, Emilia G\'{o}mez$^2$, Zheng-Hua Tan$^1$,\\ Jesper Jensen$^{1,3}$}
\address{
$^1$ Aalborg University, Aalborg, Denmark\\
$^2$ Universitat Pompeu Fabra, Barcelona, Spain \\
$^3$ Oticon A/S, Sm{\o}rum, Denmark
}
\email{\{danmi, zt, jje\}@es.aau.dk $\;$ \{olga.slizovskaia, gloria.haro, emilia.gomez\}@upf.edu}

\begin{document}

\maketitle
\begin{abstract}
Both acoustic and visual information influence human perception of speech. For this reason, the lack of audio in a video sequence determines an extremely low speech intelligibility for untrained lip readers. In this paper, we present a way to synthesise speech from the silent video of a talker using deep learning. The system learns a mapping function from raw video frames to acoustic features and reconstructs the speech with a vocoder synthesis algorithm. To improve speech reconstruction performance, our model is also trained to predict text information in a multi-task learning fashion and it is able to simultaneously reconstruct and recognise speech in real time. The results in terms of estimated speech quality and intelligibility show the effectiveness of our method, which exhibits an improvement over existing video-to-speech approaches.\end{abstract}

\begin{figure*}
	\centering
		\includegraphics[scale=.49]{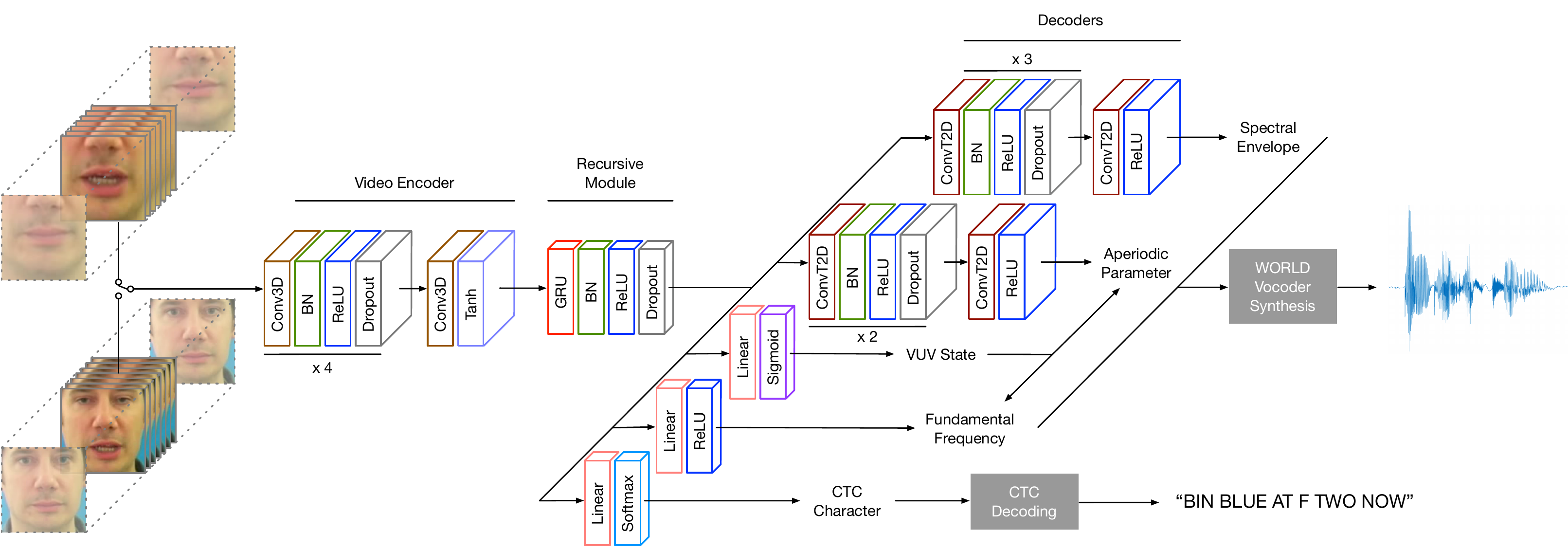}
	\caption{Pipeline of our system. Conv3D: 3-D convolution. BN: Batch normalisation. GRU: Gated recurrent unit. ConvT2D: 2-D transposed convolution. VUV:~Voiced-unvoiced. CTC: Connectionist temporal classification.}
	\label{fig:architecture}
\end{figure*}

\section{Introduction}

Most of the events that we experience in our life consist of visual and acoustic stimuli. Recordings of such events may lack the acoustic component, for example due to limitations of the recording equipment or technical issues in the transmission of the information. Since acoustic and visual modalities are often correlated, methods to reconstruct audio signals using videos have been proposed \cite{davis2014visual, owens2016visually, zhou2018visual}.

In this paper, we focus on one particular case of the aforementioned problem: \textit{speech reconstruction (or synthesis) from a silent video}. Solving this task might be useful to automatically generate speech for surveillance videos and for extremely challenging speech enhancement applications, e.g. hearing assistive devices, where noise completely dominates the target speech, making the acoustic signal worth less than its video counterpart.

A possible way to tackle the problem is to decompose it into two steps: first, a \textit{visual speech recognition} (VSR) system \cite{assael2016lipnet, Stafylakis2017, chung2017lip} predicts the spoken sentences from the video; then, a \textit{text-to-speech} (TTS) model \cite{sotelo2017char2wav, ping2018deep, shen2018natural} synthesises speech based on the output of the VSR system. However, at least two drawbacks can be identified when such an approach is used. In order to generate speech from text, each word should be spoken in its entirety to be processed by the VSR and the TTS systems, imposing great limitations for real-time applications. Furthermore, when the TTS method is applied, useful information that should be captured by the system, such as emotion and prosody, gets lost, making the synthesised speech unnatural. For these reasons, approaches that estimate speech from a video, without using text as an intermediate step, have been proposed.

Le Cornu and Miller \cite{cornu2015reconstructing, le2017generating} developed a video-to-speech method with a focus on speech intelligibility rather than quality. This is achieved by estimating spectral envelope (SP) audio features from visual features and then reconstructing the time-domain signal with the STRAIGHT vocoder \cite{kawahara1999restructuring}. Since the vocoder also requires other audio features, i.e. the fundamental frequency (F0) and the aperiodic parameter (AP), these are artificially created independently of the visual features.

Ephrat and Peleg \cite{ephrat2017vid2speech} treated speech reconstruction as a regression problem using a neural network which takes as input raw visual data and predicts a line spectrum pairs (LSP) representation of linear predictive coding (LPC) coefficients computed from the audio signal. The waveform is reconstructed from the estimated audio features using Gaussian white noise as excitation, producing unnatural speech. This issue is tackled in a subsequent work \cite{ephrat2017improved}, where a neural network estimates the mel-scale spectrogram of the audio from video frames and optical flow information derived from the visual input. The time-domain speech signal is reconstructed using either example-based synthesis, in which estimated audio features are replaced with their closest match in the training set, or speech synthesis from predicted linear-scale spectrograms.

Akbari et. al. \cite{akbari2018lip2audspec} tried to reconstruct natural sounding speech using a neural network that takes as input the face region of the talker and estimates bottleneck features extracted from the auditory spectrogram by a pre-trained autoencoder. The time-domain signal is obtained with the algorithm in \cite{chi2005multiresolution}. This approach shows its effectiveness when compared to \cite{ephrat2017vid2speech}.

All the methods reported until now have a major limitation: they estimate either a magnitude spectrogram, SPs or LSPs, which do not contain all the information of a speech signal. Vougioukas et al. \cite{Vougioukas2019} addressed this issue and proposed an end-to-end model that can directly synthesise audio waveforms from videos using a generative adversarial network (GAN). However, their direct estimation of a time-domain signal causes artefacts in the reconstructed speech.

%Furthermore, their approach is capable to reconstruct intelligible speech in a speaker independent scenario, despite a substantial performance gap with the speaker dependent case.

In this work, we propose an approach, \textit{vid2voc}, to estimate WORLD vocoder \cite{morise2016world} features from the silent video of a speaker\footnote{Although this paper aims at synthesising speech from frontal-view silent videos, it is worth mentioning that some methods using multi-view video feeds have also been developed \cite{kumar2018harnessing, kumar2018mylipper, kumar2019lipper, Uttam2019}.}. We trained the systems using either the whole face or the mouth region only, since previous work \cite{ephrat2017vid2speech} shows a benefit in using the entire face. Our method differs from the work in \cite{cornu2015reconstructing, le2017generating}, because we predict all the vocoder features (not only SP) directly from raw video frames. The estimation of F0 and AP, alongside with SP, allows to have a framework with a focus on speech intelligibility (as in \cite{cornu2015reconstructing, le2017generating}) and speech quality, able to outperform even the recently proposed GAN-based approach in \cite{Vougioukas2019} in several conditions.
In addition, we train a system that can simultaneously perform speech reconstruction (our main goal) and VSR, in a multi-task learning fashion. This can be useful in all the applications that require video captioning without adding considerable extra complexity to the system. Although Kumar et al. \cite{kumar2019lipper} incorporate a text-prediction model in their multi-view speech reconstruction pipeline, this model is trained separately from the main system and it is quite simple: it classifies encoded audio features estimated with a pre-trained network into 10 text classes. This makes the method dependent on the number of different sentences of the specific database used for training and not suitable for real-time applications. Instead, we make use of the more flexible connectionist temporal classification (CTC) \cite{graves2006connectionist} sequence modelling which has already shown its success in VSR \cite{assael2016lipnet}.

Additional material, including samples of reconstructed speech that the reader is encouraged to listen to for a better understanding of the effectiveness of our approach, can be found in \url{https://danmic.github.io/vid2voc/}.

%%%%%%%%%%%%%%%%%%%%%%%%%%%%%%%%%%%%%%%%%%%%%%%%%%%%
\section{Methodology and Experimental Setup}

\subsection{Audio-Visual Speech Corpus}

Experiments are conducted on the GRID corpus \cite{cooke2006audio}, which consists of audio and video recordings from 34 speakers (s1$-$34), 18 males and 16 females, each of them uttering 1000 six-word sentences with the following structure: $<$command$>$ $<$color$>$ $<$preposition$>$ $<$letter$>$ $<$digit$>$ $<$adverb$>$. Each video has a resolution of 720$\times$576 pixels, a duration of 3 s and a frame rate of 25 frames per second. The audio tracks have the same duration as the videos and a sample frequency of 50 kHz. In addition, text transcription for every utterance is provided.

As in \cite{Vougioukas2019}, we evaluate our systems in speaker dependent and speaker independent settings. Regarding the speaker dependent scenario, the data from 4 speakers (s1, s2, s4, s29) is pooled together, then 90\% of the data is used for training, 5\% for validation and 5\% for testing. Regarding the speaker independent scenario, the data from 15 speakers (s1, s3, s5$-$8, s10, s12, s14, s16, s17, s22, s26, s28, s32) is used for training, the data from 7 speakers (s9, s20, s23, s27, s29, s30, s34) for validation and the data from 10 speakers (s2, s4, s11, s13, s15, s18, s19, s25, s31, s33) for testing.

\subsection{Audio and Video Preprocessing}

The acoustic model used in this work is based on the WORLD vocoder \cite{morise2016world}, with a sample frequency of 50 kHz and a hop size of 250 samples\footnote{The window length is automatically determined by the WORLD algorithm.}. WORLD consists of three analysis algorithms to determine SP, F0 and AP features, and a synthesis algorithm which incorporates these three features. Here, we use SWIPE \cite{camacho2007swipe} and D4C \cite{morise2016d4c} to estimate F0 and AP, respectively. As done in \cite{blaauw2017neural}, a dimensionality reduction of the features is applied: SP is reduced to 60 log mel-frequency spectral coefficients (MFSCs) and
%obtained by truncated frequency warping in the cepstral domain \cite{tokuda1994mel} using an all-pole filter with a warping coefficient of 0.56; 
AP is reduced to 5 coefficients according to the D4C band-aperiodicity estimation. In addition, a voiced-unvoiced (VUV) state is obtained by thresholding the F0 obtained with SWIPE. All the acoustic features are min-max normalised using the statistics of the training set as in \cite{chandna2019vocoder}.

As in \cite{Vougioukas2019}, videos are preprocessed as follows: first, the faces are aligned to the canonical face\footnote{We use the face processor library in \url{https://github.com/DinoMan/face-processor}, which makes use of \cite{bulat2017far}.}; then, the video frames are normalised in the range $[-1, 1]$, resized to 128$\times$96 pixels and, for the models that use only the mouth region as input, cropped preserving the bottom half; finally, the videos are mirrored with a probability of 0.5 during training.

\subsection{Architecture and Training Procedure}

As shown in Figure \ref{fig:architecture}, our network maps video frames of a speaker to vocoder features and consists of a \textit{video encoder}, a \textit{recursive module} and five decoders: \textit{SP decoder}, \textit{AP decoder}, \textit{VUV decoder}, \textit{F0 decoder} and \textit{VSR decoder}. We also tried not to use the VSR decoder, to see whether it has any impact on the performance.

The video encoder is inspired by \cite{Vougioukas2019}: it takes as input one video frame concatenated with the three previous and the three next frames and applies five 3-D convolutions (conv3D). Each of the first four convolutional layers is followed by batch normalisation (BN) \cite{ioffe2015batch}, ReLU activation and dropout \cite{srivastava2014dropout}, while the last one is followed by Tanh activation.

To model the sequential nature of video data, a recursive module is used: it consists of a single-layer gated recurrent unit (GRU) \cite{cho2014learning}, BN, ReLU activation and dropout.

Each decoder takes the GRU features as input. For every video frame the SP decoder produces an eight-frame-long estimate $\widehat{W}_{se} \in {\rm I\!R}^{60\times8}$ of the normalised dimensionality-reduced SP, $W_{se}$, through three 2-D transposed convolutions (convT2D), each followed by BN, ReLU activation and dropout, and another convT2D followed by ReLU activation.

The VUV decoder consists of a linear layer followed by ReLU activation. A threshold of 0.2 is applied to the output obtaining $\widehat{{W}}_{vuv} \in {\rm I\!R}^{8}$, an estimate of the VUV state, $W_{vuv}$.
%Since the VUV state is useful to get correct estimates of F0 and AP, $\widehat{{W}}_{vuv}$ is used in the way that we are going to describe in the rest of the section.

The AP decoder has a structure similar to the SP decoder, with a total of three convT2D in this case. Its output, $O_{nap} \in {\rm I\!R}^{5\times8}$, together with $\widehat{{W}}_{vuv}$ is used to get $\widehat{W}_{nap}$, an estimate of $W_{nap} = I_{5,8} - W_{ap}$, where $I_{5,8}$ indicates an all-ones matrix with 5 rows and 8 columns, and $W_{ap}$ is the normalised dimensionality-reduced AP:
\begin{align}
\begin{split}\label{eqn:nap}
   (\widehat{W}_{nap})_i ={}& (O_{nap})_i \odot  \widehat{W}_{vuv} \;\;\; \text{for} \; i \in \{1, \dots, 5\}
\end{split}
\end{align}
where $(A)_i$ indicates the $i$-th row of $A$ and $\odot$ denotes the element-wise product.

The F0 decoder has a linear layer followed by a sigmoid activation function. Its output, $O_{f0} \in {\rm I\!R}^{8}$, is point-wise multiplied with $\widehat{{W}}_{vuv}$ to obtain $\widehat{{W}}_{f0}$, an estimate of the normalised F0, $W_{f0}$:
\begin{align}
\begin{split}\label{eq:f0}
   \widehat{{W}}_{f0} ={}& O_{f0} \odot  \widehat{W}_{vuv}.
\end{split}
\end{align}

Finally, the VSR decoder, consisting of a linear and a softmax layers, outputs a CTC character that will be used to predict the text transcription of the utterance. 

The system is trained to minimise the following loss:
\begin{multline}J =
\frac{{\lambda}_{1}}{{\lambda}} J_{se} + 
\frac{{\lambda}_{2}}{{\lambda}}  J_{nap} +
\frac{{\lambda}_{3}}{{\lambda}}  J_{f0} +
\frac{{\lambda}_{4}}{{\lambda}}  J_{vuv} +
\frac{{\lambda}_{5}}{{\lambda}}  J_{vsr}
\label{eqn:loss}\end{multline}
where ${\lambda}_{1}=600$, ${\lambda}_{2}=50$, ${\lambda}_{3}=10$, ${\lambda}_{4}=10$, ${\lambda}_{5}=1$, ${\lambda}=\sum_{i=1}^{5}{{\lambda}_{i}}$ and:
\begin{itemize}
	\item $J_{se}$: mean squared error (MSE) between $W_{se}$ and $\widehat{W}_{se}$.
	\item $J_{nap}$: MSE between $W_{nap}$ and $\widehat{W}_{nap}$.
	\item $J_{f0}$: MSE between ${W}_{f0}$ and $\widehat{W}_{f0}$.
	\item $J_{vuv}$: MSE between ${W}_{vuv}$ and $\widehat{W}_{vuv}$.
	\item $J_{vsr}$: CTC loss \cite{graves2006connectionist} between the target text transcription and the estimated one.
\end{itemize}

Details regarding architecture and training hyperparameters can be found in Table \ref{tab:architecture}.

\begin{table}
\caption{Architecture and training hyperparameters. Activation, batch normalisation and dropout omitted for brevity.}
\centering
\resizebox{0.45\textwidth}{!}{
\begin{tabular}{c | c c c c c }
\toprule
\multicolumn{6}{c}{\textbf{Input Size}}\\
\bottomrule
\multicolumn{6}{c}{$B{\times}S{\times}C{\times}F{\times}H{\times}W$}\\
\toprule
\multicolumn{6}{c}{\textbf{Video Encoder}}\\
\bottomrule
{Layer} & {\begin{tabular}{@{}c@{}}Input \\ Channels\end{tabular}}   & {\begin{tabular}{@{}c@{}}Output \\ Channels\end{tabular}}   & {\begin{tabular}{@{}c@{}}Kernel \\ Size\end{tabular}}   & Stride   & Padding    \\
\midrule
Conv3D &        3        &      64      &   (7,4,4)      &     (1,2,2)          &   (0,1,1)      \\ 
Conv3D &      64        &    128      &  (1,4,4)       &     (1,2,2)          &   (0,1,1)      \\ 
Conv3D &    128        &    256      &   (1,4,4)       &    (1,$d_1$,2)  &   (0,1,1)      \\ 
Conv3D &    256        &    512      &   (1,4,4)       &     (1,2,2)         &   (0,1,1)      \\ 
Conv3D &    512        &    128      &   (1,$d_2$,6) &     (1,1,1)       &   (0,0,0)      \\ 
\toprule
\multicolumn{6}{c}{\textbf{Recursive Module}}\\
\bottomrule
{Layer} & \multicolumn{2}{c}{Input Size}   & \multicolumn{2}{c}{Hidden Size}      \\
\midrule
GRU &        \multicolumn{2}{c}{128}        &      \multicolumn{2}{c}{128}               \\ 
\toprule
\multicolumn{6}{c}{\textbf{Spectral Envelope (SP) Decoder}}\\
\bottomrule
{Layer} & {\begin{tabular}{@{}c@{}}Input \\ Channels\end{tabular}}   & {\begin{tabular}{@{}c@{}}Output \\ Channels\end{tabular}}   & {\begin{tabular}{@{}c@{}}Kernel \\ Size\end{tabular}}   & Stride   & Padding   \\
\midrule
ConvT2D &    128        &    256      &   (1,6)       &     (1,1) &     (0,0)   \\ 
ConvT2D &    256        &    128      &   (2,4)       &     (1,2) &     (0,0)     \\ 
ConvT2D &    128        &      64      &   (4,4)       &      (1,2) &     (0,0)     \\ 
ConvT2D &      64        &        1      &   (4,2)       &     (1,2)  &     (0,0)    \\ 
\toprule
\multicolumn{6}{c}{\textbf{Aperiodic Parameter (AP) Decoder}}\\
\bottomrule
{Layer} & {\begin{tabular}{@{}c@{}}Input \\ Channels\end{tabular}}   & {\begin{tabular}{@{}c@{}}Output \\ Channels\end{tabular}}   & {\begin{tabular}{@{}c@{}}Kernel \\ Size\end{tabular}}   &  Stride   & Padding  \\
\midrule
ConvT2D &    128        &    128      &   (4,1)  &     (1,1) &     (0,0)         \\ 
ConvT2D &    128        &      64      &   (3,3)  &     (1,1) &     (0,0)         \\ 
ConvT2D &      64        &        1      &   (3,3)  &     (1,1) &     (0,0)          \\ 
\toprule
\multicolumn{6}{c}{\textbf{Voiced-Unvoiced (VUV) Decoder}}\\
\bottomrule
{Layer} & \multicolumn{2}{c}{Input Size}   & \multicolumn{2}{c}{Output Size}      \\
\midrule
Linear &        \multicolumn{2}{c}{128}        &      \multicolumn{2}{c}{8$^{\mathrm{a}}$}               \\ 
\toprule
\multicolumn{6}{c}{\textbf{Fundamental Frequency (F0) Decoder}}\\
\bottomrule
{Layer} & \multicolumn{2}{c}{Input Size}   & \multicolumn{2}{c}{Output Size}      \\
\midrule
Linear &        \multicolumn{2}{c}{128}        &      \multicolumn{2}{c}{8$^{\mathrm{a}}$}               \\ 
\toprule
\multicolumn{6}{c}{\textbf{Visual Speech Recognition (VSR) Decoder}}\\
\bottomrule
{Layer} & \multicolumn{2}{c}{Input Size}   & \multicolumn{2}{c}{Output Size}      \\
\midrule
Linear &        \multicolumn{2}{c}{128}        &      \multicolumn{2}{c}{28$^{\mathrm{b}}$}               \\ 
\toprule
\multicolumn{6}{c}{\textbf{Extra Information}}\\
\bottomrule
\multicolumn{6}{l}{The system is implemented in Pytorch \cite{NEURIPS2019_9015} and trained for $N$}\\
\multicolumn{6}{l}{iterations using the Adam optimizer \cite{kingma2014adam} with a learning rate}\\
\multicolumn{6}{l}{of 0.0001, $\beta_1$=0.5 and $\beta_2$=0.9. The model that performs the}\\ 
\multicolumn{6}{l}{best in terms of PESQ on the validation set is used for testing.}\\
\multicolumn{6}{l}{$S$=75 (sequence length). $C$=3 (image channels).}\\
\multicolumn{6}{l}{$F$=7 (consecutive video frames). $W$=96 (video frame width).}\\
\multicolumn{6}{l}{If the full face is used as input:}\\
\multicolumn{6}{l}{$B$=16 (batch size). $H$=128 (video frame height). $d_1$=3. $d_2$=5.}\\
\multicolumn{6}{l}{If only the mouth is used as input:}\\
\multicolumn{6}{l}{$B$=24 (batch size). $H$=64 (video frame height). $d_1$=2. $d_2$=4.}\\
\multicolumn{6}{l}{In the speaker dependent case, the dropout probability of each}\\
\multicolumn{6}{l}{dropout layer is $p_d$=0.2. $N$=300000.}\\ 
\multicolumn{6}{l}{In the speaker independent case, $p_d$=0.5 for the video encoder}\\ 
\multicolumn{6}{l}{and the GRU, and $p_d$=0.2 for the rest. $N$=185000.}\\ 
\bottomrule
\multicolumn{6}{l}{$^{\mathrm{a}}$Eight is the number of the output audio frames corresponding}\\
\multicolumn{6}{l}{to the video frame used as input (together with its context).}\\
\multicolumn{6}{l}{$^{\mathrm{b}}$The 28 CTC characters consist of the 26 letters of the English}\\ 
\multicolumn{6}{l}{alphabet, one space character and one blank token.}\\
\end{tabular}}
\label{tab:architecture}
\end{table}

\begin{table}
\caption{Systems used in this study.}
\centering
\resizebox{0.45\textwidth}{!}{%
\begin{tabular}{l c c}
\toprule
\multicolumn{1}{c}{}   & \multicolumn{2}{c}{Input}\\
\midrule
 & Mouth & Face \\
\midrule
w/o VSR Decoder & vid2voc-M & vid2voc-F  \\
w/ VSR Decoder & vid2voc-M-VSR & vid2voc-F-VSR  \\
\bottomrule
\end{tabular}}
\label{tab:systems}
\end{table}

\subsection{Waveform Reconstruction and Lipreading}

The network outputs are used to reconstruct the speech waveform with the WORLD synthesis algorithm \cite{morise2016world} and to get a text transcription adopting the best path CTC decoding scheme~\cite{graves2006connectionist}.

\subsection{Evaluation Metrics}

The system is evaluated in terms of perceptual evaluation of speech quality (PESQ) \cite{rix2001perceptual} and extended short-time objective intelligibility (ESTOI) \cite{jensen2016algorithm}, two of the most used measures that provide estimates of speech quality and speech intelligibility, respectively. PESQ scores are in the range from $-0.5$ to $4.5$ and ESTOI scores practically lie between $0$ and $1$. In both cases, higher values correspond to better performance.

For the systems having the VSR decoder, we also provide the word error rate (WER), a standard metric for automatic speech recognition systems. In this case, lower values correspond to better performance.

%%%%%%%%%%%%%%%%%%%%%%%%%%%%%%%%%%%%%%%%%%%%%%%%%%%%

\section{Results and Discussion}

As shown in Table \ref{tab:systems}, four systems are trained based on the input (mouth or full face) and the presence of the VSR decoder (only speech synthesis or speech synthesis and VSR).

The systems are compared with the recently proposed GAN-based approach in \cite{Vougioukas2019}.
%, which achieved state-of-the-art performance in almost all the objective measures used in \cite{Vougioukas2019}. 
As an additional baseline, we also report the PESQ score for \cite{akbari2018lip2audspec}, since this method, which makes use of bottleneck features extracted from auditory spectrograms,  outperforms \cite{Vougioukas2019} in terms of estimated speech quality for the speaker dependent case.

\begin{table}
\caption{Results for the speaker dependent and the speaker independent cases. Best performance (except WORLD) in bold.}
\centering
\resizebox{0.45\textwidth}{!}{%
\begin{tabular}{l | c c c | c c c}
\toprule
\multicolumn{1}{c}{}   & \multicolumn{3}{c}{Speaker Dependent} & \multicolumn{3}{c}{Speaker Independent}\\
\midrule
Mean Scores & PESQ $\uparrow$ & ESTOI $\uparrow$ & WER $\downarrow$ & PESQ $\uparrow$ & ESTOI $\uparrow$ & WER $\downarrow$  \\
\midrule
Approach in \cite{akbari2018lip2audspec}$^{\mathrm{a}}$  & 1.82 & - & - & - & - & - \\
Approach in \cite{Vougioukas2019} & 1.71 & 0.329 & - & 1.24 & 0.198 & - \\
vid2voc-M & 1.89 & 0.448 & - &  1.20 & 0.214 & - \rule{0pt}{3ex} \\
vid2voc-M-VSR & \textbf{1.90} & \textbf{0.455} & 15.1\% & 1.23 & \textbf{0.227} & \textbf{51.6\%} \\
vid2voc-F & 1.85 & 0.439 & - & 1.19 & 0.202 & -  \rule{0pt}{3ex}\\
vid2voc-F-VSR & 1.88 & 0.447 & \textbf{14.4\%} & \textbf{1.25} & 0.210 & 69.3\% \\
WORLD$^{\mathrm{b}}$ & 3.06 & 0.759 & - & 3.03 & 0.759 & - \rule{0pt}{3ex}\\
%\midrule
%Male Scores & PESQ & ESTOI & WER & PESQ & ESTOI & WER \\
%\midrule
%Approach in \cite{Vougioukas2019} & 1.96 & 0.296 & - & 1.63 & 0.253 & - \\
%
%vid2voc-M & 2.14 & 0.412 & - & 1.61 & 0.268 & - \rule{0pt}{3ex}\\
%vid2voc-M-VSR & \textbf{2.18} & \textbf{0.421} & \textbf{15.1\%} & \textbf{1.67} & \textbf{0.278} & \textbf{49.6\%} \\
%
%vid2voc-F & ? & ? & - & 1.62 & 0.250 & - \rule{0pt}{3ex}\\
%vid2voc-F-VSR & 2.17 & 0.413 & \textbf{15.1\%} & \textbf{1.67} & 0.269 &  67.2\% \\
%
%WORLD$^{\mathrm{b}}$ & 3.08 & 0.707 & - & 3.01 & 0.757 & - \rule{0pt}{3ex}\\
%\midrule
%Female Scores & PESQ & ESTOI & WER & PESQ & ESTOI & WER \\
%\midrule
%Approach in \cite{Vougioukas2019} & 1.45 & 0.362 & - & \textbf{1.07} & 0.174 & - \\
%
%vid2voc-M & \textbf{1.63} & 0.485 & - & 1.02 & 0.190 & - \rule{0pt}{3ex}\\
%vid2voc-M-VSR & 1.62 & \textbf{0.488} & 15.1\% & 1.04 & \textbf{0.205} & \textbf{52.4}\% \\
%
%vid2voc-F & \textbf{?} & ? & - & 1.00 & 0.182 & - \rule{0pt}{3ex}\\
%vid2voc-F-VSR & 1.59 & 0.481 & \textbf{13.7\%} & 1.06 & 0.185 & 70.2\% \\
%
%WORLD$^{\mathrm{b}}$ & 3.05 & 0.812 & - & 3.04 & 0.759 & - \rule{0pt}{3ex}\\
\bottomrule
\multicolumn{7}{l}{$^{\mathrm{a}}$Value taken from the experiments in \cite{Vougioukas2019}.}\\
\multicolumn{7}{l}{$^{\mathrm{b}}$WORLD indicates the reconstruction retrieved from the vocoder features of the}\\
\multicolumn{7}{l}{clean speech signals and it is a performance upper bound of our systems.}
\end{tabular}}
\label{tab:results}
\end{table}

\subsection{Speaker Dependent Case}

Table \ref{tab:results} (left part) shows the speaker dependent results. We observe that our models outperform the approach in \cite{Vougioukas2019} in terms of both PESQ and ESTOI by a considerable margin. Vougioukas et al. \cite{Vougioukas2019} mention that their  system produces low-power hum artefacts that affect the performance. They tried to solve the issue by applying average filtering to the output of their network, experiencing a rise of the PESQ score from 1.71 to 1.80 (not shown in Table \ref{tab:results}), comparable to \cite{akbari2018lip2audspec}, but still appreciably lower than the results we achieve. However, this filtering negatively affected the intelligibility of the produced speech signals, and was not used in the final system.

Among the systems we developed (cf. Table \ref{tab:systems}), we observe that including the VSR decoder in the pipeline is beneficial for the speech reconstruction task (see Table \ref{tab:results}). Moreover, the use of the mouth as input not only is sufficient to synthesise speech, but it also allows to achieve higher estimated speech quality and intelligibility if compared to the models that use the whole face of the speaker as input. This might be explained by the fact that handling an input with a larger dimensionality is harder if we want to keep roughly the same deep architecture with a similar number of parameters. However, when the whole face is used as input, the WER is slightly lower, indicating that there might be a performance trade-off between VSR and speech reconstruction that should be further investigated in future work in relation with other multi-task learning techniques.

\begin{figure}
	\centering
		\includegraphics[scale=.43]{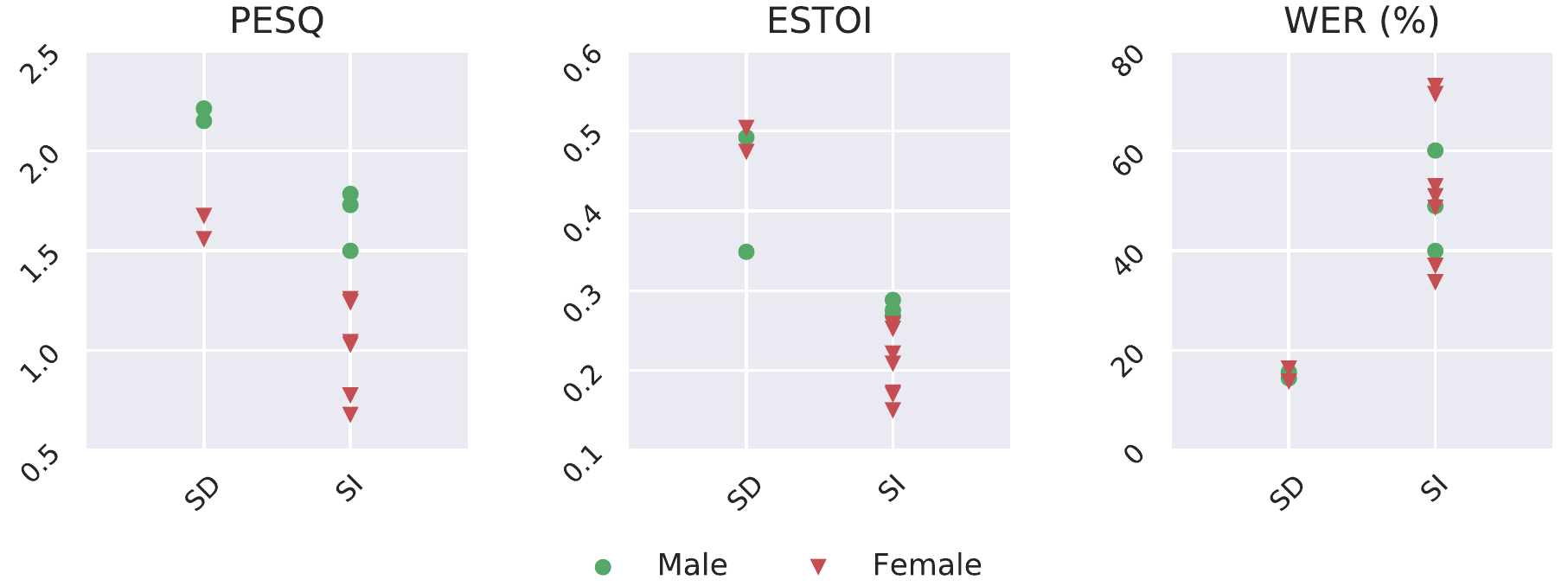}
	\caption{Results of the vid2voc-M-VSR models for the speaker dependent (SD) and the speaker independent (SI) cases. Each marker indicates the mean score of a speaker.}
	\label{fig:plot}
\end{figure}

\subsection{Speaker Independent Case}

Regarding the speaker independent scenario (cf. right part of Table \ref{tab:results}), we observe that the performance gap between the approach in \cite{Vougioukas2019} and our systems is not as large as for the speaker dependent case. Although our models appear to perform slightly better than  \cite{Vougioukas2019} in terms of ESTOI, the PESQ scores are similar. This can be explained by the fact that some speech characteristics, e.g. F0, cannot be easily estimated for unseen speakers. Since it is reasonable to think that people having similar facial characteristics (e.g. due to gender, age etc.) have similar speech characteristics (cf. \cite{oh2019speech2face}, where the face of a person was predicted from a speech signal), we expect that training a network with a dataset that includes more speakers might be beneficial: such a network can produce an average voice of speakers from the training set that share similar facial traits with an unseen talking face.

Among the systems we developed, the presence of the VSR decoder still gives an advantage for speech reconstruction. Unlike the speaker dependent case, the WER for the model that uses the whole face as input is higher than the system using only the mouth. This is due to the early stopping technique that we adopt, which tends to favour speech reconstruction over VSR, indicating again the trade-off between these two tasks.

Finally, Figure \ref{fig:plot} shows the results for the vid2voc-M-VSR models by speaker. We can see that the spread of the scores is much higher for the speaker independent case in particular for WER. This is in line with the observations reported in \cite{Vougioukas2019}, suggesting the different performance between the estimated speech of subjects whose facial traits substantially differ from the speakers in the training set and the others.

% that can be captured and exploited by a deep network, which can associate a voice to a talking face more accurately than a system trained with a limited number of speakers.

%%%%%%%%%%%%%%%%%%%%%%%%%%%%%%%%%%%%%%%%%%%%%%%%%%%%
\section{Conclusion}

In this study, we reconstructed speech from silent videos using a deep model that estimates WORLD vocoder features. We tested our approach in both speaker dependent and speaker independent scenarios. In both cases, we were able to obtain speech signals with estimated speech quality and intelligibility generally higher if compared to a recently proposed GAN-based approach. In addition, we designed our system to simultaneously perform visual speech recognition by using a decoder that estimates CTC characters from a given video sequence.

Future work includes: (a) the adoption of self-paced multi-task learning techniques; (b) the improvement of the visual speech recognition performance, e.g. with a beam search decoding scheme; (c) the design of a system that can generalise well to unseen speakers in noncontrolled environments.

%%%%%%%%%%%%%%%%%%%%%%%%%%%%%%%%%%%%%%%%%%%%%%%%%%%%
\section{Acknowledgment}

%The authors would like to thank: Konstantinos Vougioukas and Stavros Petridis, for the help regarding their video-driven speech reconstruction system; Pritish Chandna and Merlijn Blaauw, for the valuable discussions about speech synthesis with vocoder features.

The authors would like to thank Konstantinos Vougioukas, Stavros Petridis, Pritish Chandna and Merlijn Blaauw.

This research is partially funded by: the William Demant Foundation; the TROMPA H2020 project (770376); the Spanish Ministry of Economy and Competitiveness under the Mar{\'i}a de Maeztu Units of Excellence Program (MDM-2015-0502) and the Social European Funds; the MICINN/FEDER UE project (PGC2018-098625-B-I00); the H2020-MSCA-RISE-2017 project (777826 NoMADS).

\bibliographystyle{IEEEtran}

\bibliography{vid2voc}

% \begin{thebibliography}{9}
% \bibitem[1]{Davis80-COP}
%   S.\ B.\ Davis and P.\ Mermelstein,
%   ``Comparison of parametric representation for monosyllabic word recognition in continuously spoken sentences,''
%   \textit{IEEE Transactions on Acoustics, Speech and Signal Processing}, vol.~28, no.~4, pp.~357--366, 1980.
% \bibitem[2]{Rabiner89-ATO}
%   L.\ R.\ Rabiner,
%   ``A tutorial on hidden Markov models and selected applications in speech recognition,''
%   \textit{Proceedings of the IEEE}, vol.~77, no.~2, pp.~257-286, 1989.
% \bibitem[3]{Hastie09-TEO}
%   T.\ Hastie, R.\ Tibshirani, and J.\ Friedman,
%   \textit{The Elements of Statistical Learning -- Data Mining, Inference, and Prediction}.
%   New York: Springer, 2009.
% \bibitem[4]{YourName17-XXX}
%   F.\ Lastname1, F.\ Lastname2, and F.\ Lastname3,
%   ``Title of your INTERSPEECH 2020 publication,''
%   in \textit{Interspeech 2020 -- 20\textsuperscript{th} Annual Conference of the International Speech Communication Association, September 15-19, Graz, Austria, Proceedings, Proceedings}, 2020, pp.~100--104.
% \end{thebibliography}

\end{document}